\documentclass{rnaastex}

\begin{document}

\title{Parallactic distances and proper motions of virtually all stars in the $\sigma$ Orionis cluster or: How I learned to get the most out of {\tt TOPCAT} and love {\em Gaia} DR2}

\correspondingauthor{Jos\'e A. Caballero}
\email{\tt caballero@cab.inta-csic.es}

\author[0000-0002-7349-1387]{Jos\'e A. Caballero}
\affiliation{Centro de Astrobiolog\'ia (CSIC-INTA),
ESAC, camino bajo del castillo s/n,
E-28692 Villanueva de la Ca\~nada, Madrid, Spain}


\section{} 

The young $\sigma$~Orionis open cluster is a cornerstone for studying X-ray emission, discs, jets, accretion, photometric variability, magnetism, abundances, and mass function of OB-type, Herbig Ae/Be and T Tauri stars, brown dwarfs, and objects beyond the deuterium burning mass limit \citep{1967PASP...79..433G,2008A&A...478..667C}.

The {\em Gaia} Data Release 2 (DR2) was made public at noon (CEST) on 25 April 2018.
Among many other things, it contains five-parameter solutions ($\alpha$, $\delta$, $\mu_\alpha \cos{\delta}$, $\mu_\delta$, $\pi$) for 1.332 million sources \citep{2018arXiv180409365G}.
By the evening of that date, and with the help of the Virtual Observatory tools {\tt TOPCAT} and {\tt Aladin} \citep{2000A&AS..143...33B,2005ASPC..347...29T}, I had compiled more useful astro-photometric data on $\sigma$~Orionis stars than all observations carried out in the previous five decades. 
Here I share useful new data that will facilitate further studies in the cluster.

First, I downloaded all {\em Gaia} DR2 data available on the 10,742 optical sources at less than 30\,arcmin to the eponymous $\sigma$~Ori star. 
It is located exactly at the center of the cluster, which is composed by a dense core that extends from the center to $\sim$20\,arcmin and a rarified halo at larger separations \citep{2008MNRAS.383..375C}.
Because of abnormal {\em Gaia} $B_P-R_P$ colors and expected $G-K_s$ colors of known cluster low-mass stars and brown dwarfs with unequivocal features of extreme youth, I~downloaded the 2MASS \citep{2006AJ....131.1163S} data of 5,742 near-infrared sources in the survey area. 
A total of 5,524 (96.2\,\%) 2MASS sources have a {\em Gaia} counterpart at less than 1.0\,arcsec.
I~also cross-matched all these sources with the Mayrit catalog of candidate and confirmed stars and brown dwarfs in $\sigma$~Orionis \citep[cf.][]{2008A&A...478..667C}. 
In spite of being over ten years old and suffering from known flaws (see corrections in \citealt{2017arXiv170700436C}), it is still the most precise and comprehensible catalog of stars and massive brown dwarfs in the cluster. 

\begin{figure}[]
\begin{center}
\includegraphics[width=0.85\textwidth]{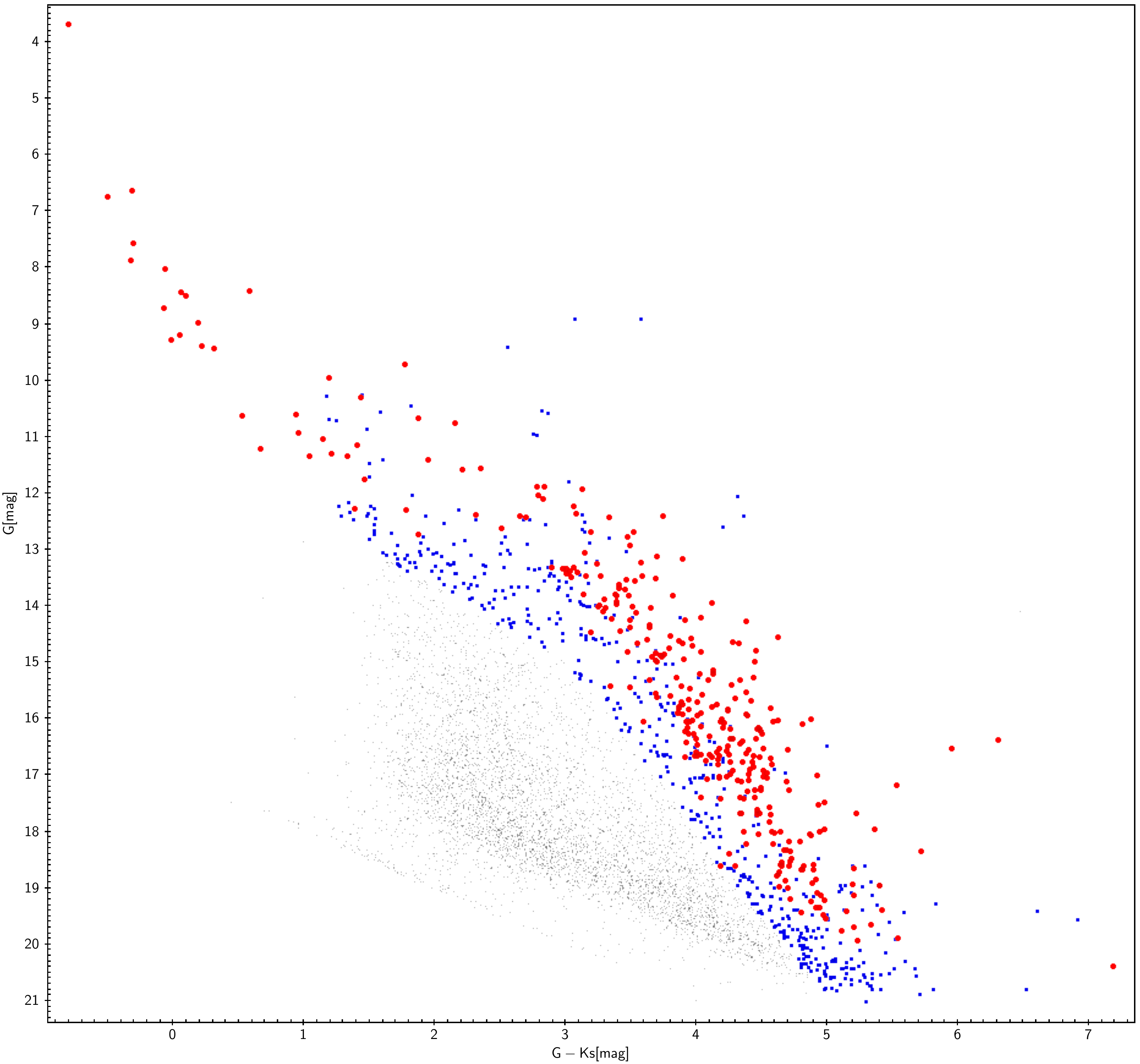}
\caption{Color-magntitude diagram $G$ vs. $G-K_s$ of all 2MASS-{\em Gaia} DR2 sources in a 30\,arcmin-radius circular area centered on the eponymous $\sigma$~Ori star, which illuminates the mane of Horsehead Nebula.
Red circles: $\sigma$~Orionis cluster member candidates in the Mayrit catalog \citep{2008A&A...478..667C};
blue dots: other sources that conservatively follow the cluster sequence (this work);
grey tiny dots: other sources in the cluster fore- and background.
The cluster sequence goes from O9.5 ($M \approx$ 17\,M$_\odot$) to M6--7 below the hydrogen burning limit ($M \approx$ 0.06--0.05\,M$_\odot$), i.e. covers almost three orders of magnitude in mass and over six orders of magnitude in optical flux. 
The virtual observatory-compliant {\tt .csv} file for reproducing this figure, or investigating by yourself membership of stars in $\sigma$~Orionis is available at RNAAS and at {\tt http://exoterrae.eu/pub/soT.csv}. 
\label{fig:1}}
\end{center}
\end{figure}

With {\tt TOPCAT}, I plotted the color-magntitude diagram $G$ vs. $G-K_s$ shown in Fig.~\ref{fig:1}.
Next, I sketched two straight lines (not shown), redward of which all $\sigma$~Orionis cluster members and candidates in the Mayrit catalog lie except for the background source 2MASS J05390697--0212168.
The 816 {\em Gaia} DR2-2MASS sources redward of the lines are available in a comma-separated values file available as data behind Figure~\ref{fig:1}. 
They consist of true cluster members and stars that conservatively follow the photometric cluster sequence but are located closer or further or do not share the very low cluster proper motion ($\sigma$~Orionis is in the antapex).

Of the 816 {\em Gaia} DR2-2MASS sources, 329 were tabulated in the Mayrit catalog (i.e. all except eight).
Of the Mayrit stars and brown dwarfs, 281 have {\em Gaia} DR2 total proper motions $\mu <$ 15\,mas\,a$^{-1}$ and parallaxes 1.5\,mas $< \pi <$ 3.7\,mas and remain as good cluster member candidates.
The 48 new outliers are stars either in the far background
(43 stars including StHa~50) 
or near foreground
(three stars: S~Ori J053724.5--021856, 
HD 294297, 
S Ori 20), 
or that are well characterized T~Tauri stars in the cluster but their parallaxes are affected by close multiplicity (two stars: Mayrit\,873229\,AB, Mayrit~359179\,AB; \citealt{2017arXiv170700436C}). 
The mean and standard deviation of the parallaxes of the resulting 281 {\em Gaia} DR2-2MASS-Mayrit stars and brown dwarfs are $\overline{\pi}$ = {\bf 2.56}\,mas and $\sigma_\pi$ = 0.29\,mas, from where one can derive $d$ = 391$\pm$44\,pc, identical within generous uncertainties to the value of 387.5$\pm$1.3\,pc derived by \citet{2016AJ....152..213S} with interferometry.

\acknowledgments

Project AYA2016-79425-C3-2-P funded by MINECO/FEDER.

\newpage

\end{document}